\documentclass[journal=inoraj,manuscript=article]{achemso}

\usepackage{chemformula} 
\usepackage[T1]{fontenc} 
\usepackage{graphicx}
\usepackage{dcolumn}
\usepackage{bm}
\usepackage{xfrac}
\usepackage{textcomp}
\usepackage[version=4]{mhchem} 
\usepackage{times}
\usepackage{kantlipsum}

\SectionNumbersOn{}
 
\emergencystretch=\maxdimen
\newcommand{\linmc}{\ce{LiNi_{0.8}Mn_{0.1}Co_{0.1}O_2}}
\newcommand{\linmcoffstoic}{\ce{Li_{1-x}Ni_{0.9+x-y}Mn_yCo_{0.1}O_2}}
\newcommand{\lno}{\ce{LiNiO_2}}
\newcommand{\lnooffstoic}{\ce{Li_{1-x}Ni_{1+x}O_2}}

\author{Paromita Mukherjee}
\email{pm545@cam.ac.uk}
\affiliation{Department of Physics, University of Cambridge, JJ Thomson Avenue, Cambridge CB3 0HE, United Kingdom}
\alsoaffiliation{The Faraday Institution, Quad One, Harwell Science and Innovation Campus, Didcot OX11 0RA, United Kingdom}
\author{Joseph A. M. Paddison}
\affiliation{Department of Physics, University of Cambridge, JJ Thomson Avenue, Cambridge CB3 0HE, United Kingdom}
\alsoaffiliation{Churchill College, University of Cambridge, Storey's Way, Cambridge CB3 0DS, United Kingdom}
\alsoaffiliation{Materials Science and Technology Division, Oak Ridge National Laboratory, Oak Ridge, TN 37831, USA}
\author{Chao Xu}
\affiliation{Department of Chemistry, University of Cambridge, Lensfield Road, Cambridge CB2 1EW, United Kingdom}
\alsoaffiliation{The Faraday Institution, Quad One, Harwell Science and Innovation Campus, Didcot OX11 0RA, United Kingdom}
\author{Zachary Ruff}
\affiliation{Department of Chemistry, University of Cambridge, Lensfield Road, Cambridge CB2 1EW, United Kingdom}
\alsoaffiliation{The Faraday Institution, Quad One, Harwell Science and Innovation Campus, Didcot OX11 0RA, United Kingdom}
\author{Andrew R. Wildes}
\affiliation{Institut Laue-Langevin, CS 20156, 38042, Grenoble Cedex 9, France}
\author{David A. Keen}
\affiliation{ISIS Neutron and Muon Source, Rutherford Appleton Laboratory, Harwell Campus,
Didcot OX11 0QX, United Kingdom}
\author{Ronald I. Smith}
\affiliation{ISIS Neutron and Muon Source, Rutherford Appleton Laboratory, Harwell Campus,
Didcot OX11 0QX, United Kingdom}
\author{Clare P. Grey}
\affiliation{Department of Chemistry, University of Cambridge, Lensfield Road, Cambridge CB2 1EW, United Kingdom}
\alsoaffiliation{The Faraday Institution, Quad One, Harwell Science and Innovation Campus, Didcot OX11 0RA, United Kingdom}
\author{Si\^an E. Dutton}
\email{sed33@cam.ac.uk}
\affiliation{Department of Physics, University of Cambridge, JJ Thomson Avenue, Cambridge CB3 0HE, United Kingdom}
\alsoaffiliation{The Faraday Institution, Quad One, Harwell Science and Innovation Campus, Didcot OX11 0RA, United Kingdom}
\title{Sample Dependence of Magnetism in the Next Generation Cathode Material \linmc{}}

\begin{document}

\makeatletter
\setlength\acs@tocentry@height{4.75cm}
\setlength\acs@tocentry@width{8.5cm}
\makeatother



\begin{abstract}
 We present a structural and magnetic study on two batches of polycrystalline \linmc{} (commonly known as Li\,NMC\,811), a Ni-rich Li ion battery cathode material, using elemental analysis, X-ray and neutron diffraction, magnetometry, and polarised neutron scattering measurements. We find that the samples, labelled S1 and S2, have the composition \linmcoffstoic{}, with $x = 0.025(2)$, $y = 0.120(2)$ for S1 and $x = 0.002(2)$, $y = 0.094(2)$ for S2, corresponding to different concentrations of magnetic ions and excess \ce{Ni^{2+}} in the \ce{Li^+} layers. Both samples show a peak in the zero-field cooled (ZFC) dc susceptibility at 8.0(2)\,K but the temperature at which the ZFC and FC (field-cooled) curves deviate is substantially different: 64(2)\,K for S1 and 122(2)\,K for S2. Ac susceptibility measurements show that the transition for S1 shifts with frequency whereas no such shift is observed for S2 within the resolution of our measurements. 
Our results demonstrate the sample dependence of magnetic properties in Li\,NMC\,811, consistent with previous reports on the parent material \lno{}. We further establish that a combination of experimental techniques are necessary to accurately determine the chemical composition of next generation battery materials with multiple cations. 
\end{abstract}

\section{Introduction}
\begin{figure}
\centering
\includegraphics[width=0.7\textwidth]{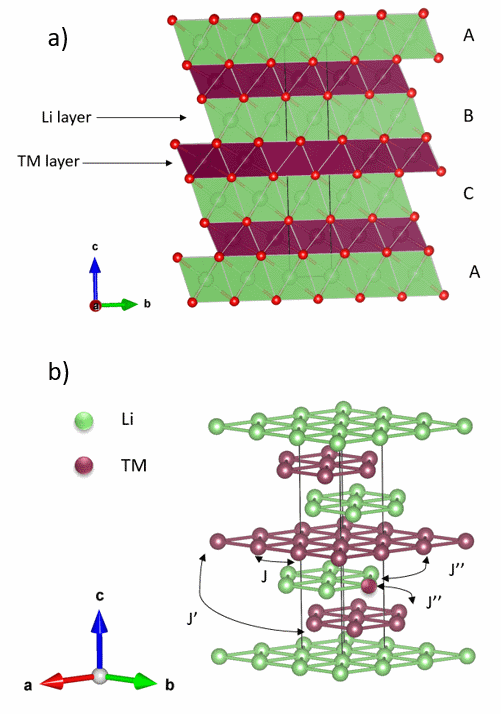}
\caption{\label{fig:Fig1} a) Crystal structure of layered Li TM oxides, showing O as red spheres, TM polyhedra in burgundy, and Li polyhedra in green. b) Competing magnetic interactions $J$, $J^{\prime}$, and $J^{\prime\prime}$ in \lno{} and Ni-rich Li TM oxides.}
\end{figure}

\lno{}, a layered transition metal (TM) oxide with $S = \sfrac{1}{2}$ \ce{Ni^{3+}} ions on a triangular lattice [Fig.~\ref{fig:Fig1}(a)],  has been widely investigated as a quantum spin liquid candidate.  However, the nature of its magnetic ground state remains controversial after decades of investigation \cite{Hirakawa1985, Hirakawa1990,Rosenberg1994,Kitaoka1998,Reynaud2001,Clancy2006}. This is because \lno{} is extremely prone to off-stoichiometry and excess of \ce{Ni^{2+}} in the \ce{Li^+} layers. Studies have shown that it is not possible to synthesise perfectly stoichiometric \lno{}, and instead the formula is \lnooffstoic{} with $x \approx 0.004$ in the best quality samples \cite{Reimers1993,Rougier1996a,Yamaura1996,Bianchi2001,Sow2012}. This results in 2$x$ $S = 1$ \ce{Ni^{2+}} spins in order to maintain charge balance (2 \ce{Ni^{2+}} = \ce{Li^+} \ce{Ni^{3+}}). Additionally, due to the similarity in \ce{Li^+} and \ce{Ni^{2+}} radii, $x$ \ce{Ni^{2+}} are in the \ce{Li^+} layers. These factors have two crucial consequences for the magnetism. First, the presence of different magnetic species ($S = \sfrac{1}{2}$ and $S = 1$) in amounts dependent on the degree of off-stoichiometry results in the magnetic ground state being highly sample-dependent. Second, the excess \ce{Ni^{2+}} in the \ce{Li^+} layers changes the competing interactions: in addition to the intra-layer ($J$) and inter-layer ($J^\prime$) interactions, there are interactions between the spins in the \ce{Li^+} layers and those in the TM layers ($J^{\prime\prime}$) [Fig.~\ref{fig:Fig1}(b)]. This may cause the spins to order magnetically or freeze instead of remaining in a dynamic liquid-like state \cite{Clancy2006,Chappel2002}. Previous  studies on \lnooffstoic{} have shown that the magnetic ground state varies significantly with slight changes in the off-stoichiometry $x$ \cite{Hirano1995,Bianchi2001,Chatterji2005,Clancy2006,Sugiyama2010,Sow2013}.

\lno{} has also been investigated as a Li ion battery cathode material as it is cheaper and less toxic than the commercially established \ce{LiCoO_2} \cite{Dahn1990,Delmas1997,Li2018,Bianchini2019}. However, safety issues due to thermal instability severely limit its practical applicability \cite{Macneil2002}. Additionally the inherent off-stoichiometry and migration of \ce{Ni^{2+}} to the \ce{Li^+} layers leads to irreversible capacity loss on long-term cycling \cite{Bianchini2019,Liu2015}. Therefore the focus has shifted towards Ni-rich compositions from the family of Li ion TM oxides with the general formula \ce{LiNi_xMn_yCo_zO_2}, $x + y + z = 1$, commonly known as Li\,NMC oxides \cite{Liu2015,Manthiram2017,Xu2017}. Being Ni-rich, these materials inherit the tendency for off-stoichiometry and \ce{Ni^{2+}} excess in \ce{Li^+} layers associated with \lno{}, which can significantly affect their performance in batteries \cite{Yin2020}.

In this paper, we focus on the next-generation Li ion battery cathode material \linmc{}, commonly known as Li\,NMC\,811. We present an investigation of the magnetic properties, crystal structure and chemical composition of two samples of Li\,NMC\,811. We find that the magnetic properties are sample-dependent, consistent with previous reports on \lno{} \cite{Reimers1993,Yamaura1996,Rougier1996,Barton2013}.
However, \linmc{} is a complex
material with multiple transition metal ions and so we cannot directly draw a one-to-one correspondence between the magnetic properties and the chemical composition. Instead, a combination of experimental techniques including elemental analysis, powder X-ray and neutron diffraction, bulk magnetic measurements, and polarised neutron diffraction are required to understand the structure and magnetism. Further, systematic studies on \lnooffstoic{} have shown that samples with relatively high levels of off-stoichiometry, $x \approx 0.04 - 0.06$, are more straightforward to identify by the non-Curie-Weiss behavior of their magnetic susceptibility and their markedly worse cycling performance. By contrast, effects of lower levels of off-stoichiometry are subtle and more varied \cite{Rougier1996,Rougier1996a,Yamaura1996,Bianchini2019}. We find the both samples studied here are close to ideal stoichiometry, but S2 is significantly closer to ideal stoichiometry than S1. This result is important because it demonstrates that the combination of techniques we employ (X-ray + neutron diffraction, bulk magnetic susceptibility, and elemental analysis) significantly increases the overall experimental sensitivity, allowing small levels of off-stoichiometry to be accurately quantified, and hence enabling more effective benchmarking of the highest quality samples.

\section{Experimental Methods}

Two batches of polycrystalline samples of Li\,NMC\,811 were obtained from Targray, each being purchased on a separate occasion. The samples will be referred to as S1 (Targray, Batch 1) and S2 (Targray, Batch 2) throughout this manuscript. Since Li\,NMC\,811 is known to be sensitive to moisture \cite{Liu2015, Dong2018}, the samples were stored in an Ar-atmosphere (\ce{O_2} $<$ 0.5 ppm, \ce{H_2O} $<$ 0.5 ppm) glovebox. All subsequent sample handling was in an Ar-atmosphere glovebox unless otherwise noted.

Elemental analysis was performed using inductively-coupled plasma optical emission spectroscopy (ICP-OES, Thermoscientific 7400 Duo). Aqueous solutions were prepared by dissolving $\sim$10\,mg of the as-received Li\,NMC\,811 powder in 1\,ml of freshly prepared, concentrated aqua regia ($3:1$ hydrochloric to nitric acid, trace element grade, Fisher Scientific) overnight and subsequently diluting with deionized water (Millipore) to $\sim$1-10\,ppm by mass for the measurement.  The concentration of a given element in the solutions was determined by comparing the emission of the sample solutions to a calibration line generated from a concentration series using a multielemental standard (VWR, Aristar\textregistered) at each wavelength of interest.  The emission wavelengths were selected such that there was no interference from other measured elements, elements in the standard or the matrix solution (2\% nitric acid). The transition metals each had multiple wavelengths which were suitable for the ICP-OES measurement and the results at each wavelength were averaged to obtain the molar value of ions in solution.  The composition of the powders was calculated by assuming that the molar fraction of transition metals was 1. The Li concentration was calculated by dividing the total moles of Li by the total moles of transition metals.

Room temperature powder X-Ray diffraction (PXRD) scans for  structural analysis were collected over $5^\circ \leq 2\theta \leq 150^\circ$ ($\Delta2\theta =0.004^\circ$) using the I11 beamline at Diamond Light Source ($\lambda = 0.826$\,\AA). Room temperature powder neutron diffraction (PND) experiments for structural characterisation were carried out on the GEM diffractometer, ISIS Neutron and Muon Source, Rutherford Appleton Laboratory, United Kingdom. The absorption correction for the time-of-flight (TOF) PND data was carried out using the Mantid program \cite{Arnold2014} and cross-checked using the GudrunN software \cite{Soper2011}. The structural Rietveld refinements \cite{Rietveld1969} were carried out using the Fullprof suite of programs \cite{Rodriguez-Carvajal1993}. The background was modelled using a Chebyshev polynomial and the peak shape was modelled using a pseudo-Voigt function for the PXRD data and an Ikeda-Carpenter function for the TOF PND data.

Magnetic dc susceptibility measurements were performed on a Quantum Design Magnetic Properties Measurement System (MPMS) with a Superconducting Quantum Interference Device (SQUID) magnetometer. The zero-field cooled (ZFC) and field-cooled (FC) susceptibility $\chi(T)$ was measured in a field of 100\,Oe in the temperature range 2-300\,K to investigate the presence of magnetic ordering. ZFC measurements were also carried out at 1000\,Oe in the same temperature range to perform Curie-Weiss fits. In a field of 1000\,Oe, the isothermal magnetisation $M(H)$ curve is linear at all $T$ and so $\chi(T)$ can be approximated by the linear relation $\chi(T)\approx M/H$. Isothermal magnetisation measurements in the field range $\mu_0H = 0$-$9$\,T for selected temperatures were carried out using the ACMS (AC Measurement System) option on a Quantum Design Physical Properties Measurement System (PPMS). ZFC ac susceptibility measurements in the temperature range 2-60\,K were carried out using the same PPMS option with a dc field of 20\,Oe and a driving field of 3\,Oe at frequencies between 1-10\,kHz.

Polarised neutron diffraction measurements for the S2 sample were carried out on the D7 diffractometer at the Institut Laue-Langevin, France, with $\lambda \sim 4.8$\,\AA. A sample of mass 7\,g was loaded in an annular Al can of diameter 20\,mm with an 18\,mm  cylindrical insert to minimise the effect of absorption from Li in our natural abundance samples. Scans were collected at 1.5\,K and 20\,K for 10 hours each. The data was processed in LAMP \cite{Richard1996} to separate the nuclear coherent, nuclear-spin incoherent, and magnetic scattering contributions.

Li NMC 811 electrodes for electrochemical characterisation of S1 and S2 were prepared by mixing 90 wt\% Li NMC 811 powder, 5 wt\% polyvinylidene difluoride (PVDF) binder, 5 wt\% carbon black (Timcal SuperP Li) and a desired amount of the NMP (1-Methyl-2-pyrrolidinone, anhydrous, 99.5\%, Sigma-Aldrich) solvent in a Thinky planetary mixer at 2000 rpm for 10 minutes in total (5 minutes per cycle and two cycles). The slurry was coated onto an Al foil and pre-dried at 100 $^\circ$C for 1 hour in a dry-room ($\sim$ -55 $^\circ$C dew point). Dried Li NMC 811 laminates were punched into circular disks with a diameter of 14 mm, which were further dried at 120 $^\circ$C for 12 hours under dynamic vacuum ($\sim 10^{-2}$ mbar) in a B\"{u}chi oven. 
2032 coin cells (Cambridge Energy Solution) were assembled in an Ar-atmosphere (\ce{O_2} $<$ 1 ppm, \ce{H_2O} $<$ 1 ppm) glovebox consisting of a 14 mm diameter Li NMC 811 cathode, 16 mm diameter Celgard 3501 separator and 15 mm diameter Li metal. 50 $\mu$L electrolyte (1 M \ce{LiPF_6}, ethylene carbonate (EC)/ethyl methyl carbonate (EMC) 3/7, SoulBrain MI) was added to each coin cell. Battery cyclings were conducted on an Arbin battery cycler at room temperature between 3.0 V and 4.4 V at various rates. C-rates were defined based on a reversible capacity of 200 mAh g$^{-1}$, for instance, for the C/5 rate (5 hours for one charge or discharge process), a current density of 40 mA g$^{-1}$ was applied.

\section{Results}

\subsection{Elemental analysis}

The composition of both samples as determined from ICP-OES is given in Table \ref{tab:Table1} and the details of the error analysis are given in the supplementary information. Both S1 and S2 have the ideal transition metal (TM) stoichiometry within error and the TM are in the expected $8:1:1$ ratio. Sample S1 is found to have a slight Li excess as compared to the nominal stoichiometry.

\begin{table}
\caption{\label{tab:Table1}Composition from ICP-OES for samples S1 and S2.}
\begin{tabular}{ccc}
\hline
Element &S1 &S2\\
\hline
Li &1.06(5)	&1.01(3)\\
Ni	&0.80(2)	&0.80(2)\\
Mn	&0.10(2)	&0.10(2)\\
Co 	&0.10(2)	&0.10(2)\\
\hline
\end{tabular}
\end{table}

\subsection{Room temperature PXRD and PND}

The room temperature PXRD data for S1 and S2 indicated that the samples were phase pure and adopt the crystal structure of Li\,NMC oxides (space group $R\bar{3}m$) with no indication of lowering of symmetry \cite{Yin2020}. The room temperature PND data were consistent with this; however, closer inspection of the data  plotted on a logarithmic intensity scale indicated a very small \ce{Li_2CO_3} impurity peak for S1, while no such peak was visible for S2. This is consistent with the higher Li content seen in elemental analysis for S1.  The amount of \ce{Li_2CO_3} for S1 from the refinement was found to be 0.2(3)\,wt\% and hence was not considered in further structural analysis.

The crystal structure was refined using a combined Rietveld refinement with the room temperature PXRD and TOF PND data using a structural model based on \lno{}, space group $R\bar{3}m$ \cite{Pouillerie2001}. PXRD is sensitive to the TMs (mainly Ni because of its higher concentration) while PND is much more sensitive to the contrast between Li (coherent scattering length $b = -1.9$\,fm) and Ni ($b = 10.3 $\,fm) as well as Mn ($b = -3.7$\,fm) and O ($b = 5.8$\,fm) \cite{Varley1992}, so a combined PND+PXRD refinement gives accurate information about the crystal structure. While carrying out the refinement, the weighting of the PXRD data was adjusted to satisfy the following two conditions simultaneously: a) the total weighting of the PXRD and 5-bank PND refinements summed to 1 (each PND bank was assigned the same weighting); b) the weighted residual of the PXRD refinement was equal to that of the PND refinement, such that the PXRD and PND data contributed equally to the refinement. The refinement of the chemical composition is discussed later in Section \ref{sec:chemcomp}.

\begin{figure}
\centering
\includegraphics[width=0.7\textwidth]{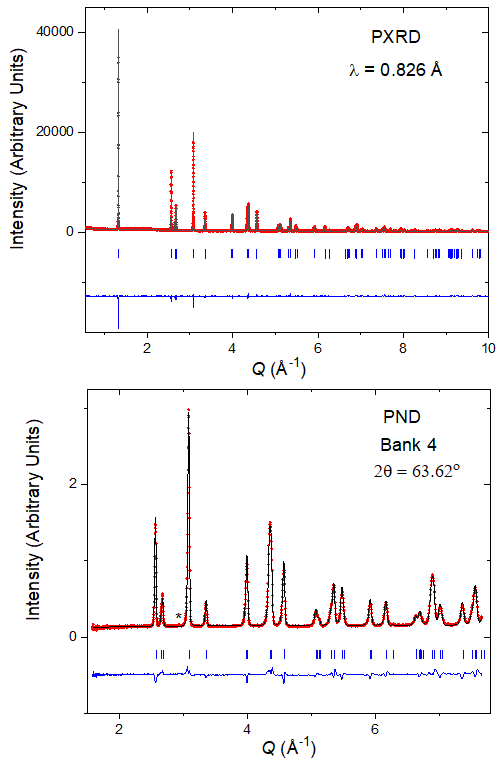}
\caption{\label{fig:Fig2} Room temperature PXRD + PND Rietveld refinement for S2. PXRD data (upper panel) were collected on I11, Diamond, and PND data (lower panel) were collected on GEM, ISIS. Data are shown as red points, fits as black lines, and difference (data--fit) as blue lines. The peak marked with * in the PND data is from the vanadium sample holder.}
\end{figure}

\begin{table}
\caption{\label{tab:Table2}Structural parameters for samples S1 and S2. All refinements were carried out in the space group $R\bar{3}m$, with Li on the $3a$ sites (0,0,0), TM (Ni, Mn, Co) on the $3b$ sites (0,0,0.5), and O on the $6c$ (0,0,$z$) sites. The \ce{Mn^{4+}} composition $y$ and \ce{Ni^{2+}} excess on the \ce{Li^+} site $x$ were allowed to vary subject to the constraints discussed in the text.}
\begin{tabular}{ccc}
\hline
Parameter &S1 &S2\\
\hline
$a$\,(\AA)	&2.8719(2)	&2.8727(3)\\
$c$\,(\AA)	&14.199(2)	&14.207(3)\\
$c/a$	&4.944(3)	&4.946(2)\\
$x$\textsuperscript{\emph{a}} 	&0.025(2)	&0.002(2)\\
$y$\textsuperscript{\emph{a}}  &0.120(2) &0.094(2) \\
$z$	&0.24095(9)	&0.24103(8)\\
$\chi^2$ &5.3	&5.7\\
\hline
$B_\textrm{iso}$\,(\AA$^2$) & & \\
\hline
Li/Ni (0,0,0)	&0.82(16)	&0.79(9)\\
TM (Ni/Mn/Co) (0,0,0.5)	&0.28(3)	&0.20(2)\\
O (0,0,$z$)	&0.68(4)	&0.66(2)\\
\hline
\end{tabular}

\textsuperscript{\emph{a}} in \linmcoffstoic{}
\end{table}

Representative fits to the PXRD and TOF PND data are shown in Figure \ref{fig:Fig2} and the refined structural parameters are compiled in Table \ref{tab:Table2}. Li-deficient \ce{LiNiO_2} with the formula \lnooffstoic{}, $x > 0.38$, crystallises in a cubic rock salt structure ($c/a = 2\sqrt{6} = 4.899$) with Li/Ni disordered on the $4a$ site; however, as the quantity of Li increases, it transforms to a hexagonal structure consisting of alternating  layers of \ce{LiO_6} and \ce{NiO_6} octahedra  and the $c/a$ value increases with the layering of the material \cite{Bianchini2019}. The $c/a$ ratio for our samples of Li\,NMC\,811 (4.944(3) for S1 and 4.946(2) for S2) is consistent with a layered hexagonal structure. Though less than that of a well-layered compound like \ce{LiCoO_2} ($c/a = 4.99$) \cite{Ohzuku1993,Akimoto1998}, it is consistent with the typical values for the parent compound \lno{} ($c/a = 4.93$) \cite{Li2018,Ohzuku1993}.  

\subsection{Bulk magnetic measurements}

\begin{figure}
\centering
\includegraphics[width=0.7\textwidth]{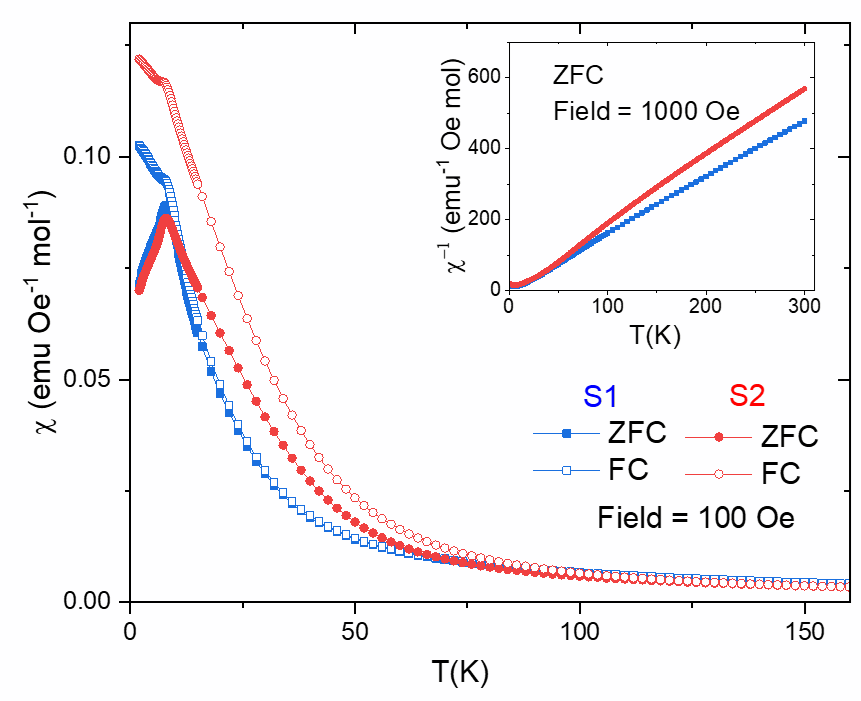}
\caption{\label{fig:Fig3} ZFC-FC dc susceptibility $\chi(T)$ in a field of 100\,Oe for S1 (blue squares) and S2 (red circles). Inset: reciprocal susceptibility $\chi^{-1}(T)$ in a field of 1000\,Oe.}
\end{figure}

\begin{figure}
\centering
\includegraphics[width=0.7\textwidth]{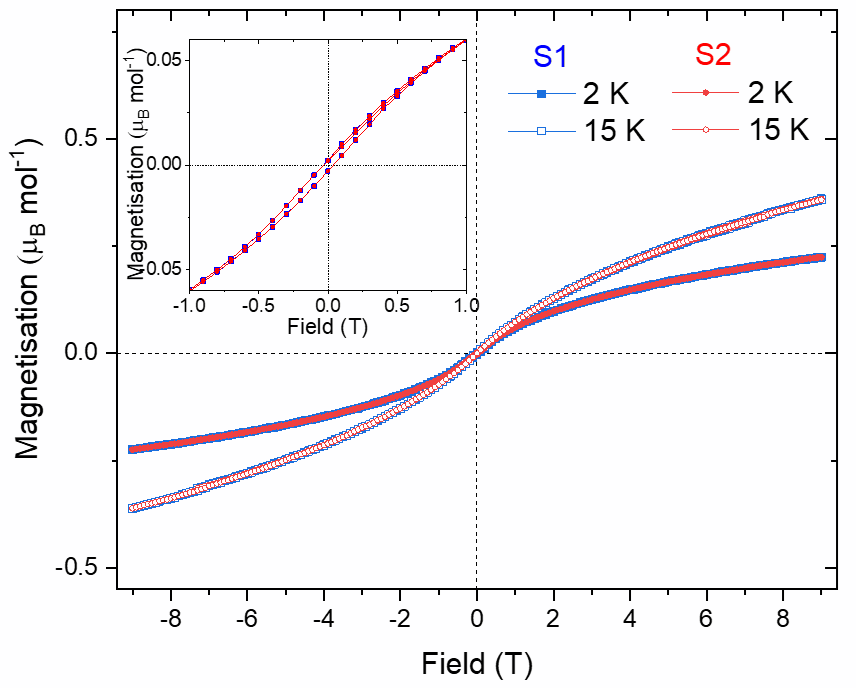}
\caption{\label{fig:Fig4} Isothermal magnetisation $M(H)$ at 2\,K and 15\,K for S1 (blue squares) and S2 (red circles). Inset: $M(H)$ at 2\,K showing the slight hysteresis at 2\,K more clearly.}
\end{figure}

Dc susceptibility $\chi(T)$ and isothermal magnetisation $M(H)$ measurements for S1 and S2 are shown in Figures~\ref{fig:Fig3} and \ref{fig:Fig4}, respectively. At low temperatures, both samples show a peak in the ZFC dc susceptibility at $T_g = 8.0(2)$\,K [Fig.~\ref{fig:Fig3}] and deviation in the ZFC-FC curves, indicating glassy behaviour. However, the temperature $T_\textrm{ZFC-FC}$ given by the temperature at which $\Delta(\chi_\textrm{FC} - \chi_\textrm{ZFC}) \approx 2$\% is very different: 64(2)\,K for S1 and 122(2)\,K for S2. The isothermal magnetisation at $T = 2$\,K ($T < T_g$) shows a slight hysteresis for both samples, consistent with a disordered ground state, whereas no hysteresis is observed at $T = 15$\,K ($T > T_g$) [Fig.~\ref{fig:Fig4}]. These features are identical for both samples within the resolution of our measurements. The $M(H)$ curves are non-linear at both $T = 2$\,K and $T = 15$\,K due to the presence of spin correlations.

\begin{table}
\caption{\label{tab:Table3}Magnetic properties for both batches of Li\,NMC\,811. Values reported for the Li\,NMC\,811\,sample in Ref.\cite{Wikberg2010} are quoted for comparison. Curie-Weiss fits were carried out in the temperature range 200-300\,K in an applied field of 1000\,Oe. Assuming ideal stoichiometry \ce{LiNi^{2+}_{0.1}Ni^{3+}_{0.7}Mn^{4+}_{0.1}Co^{3+}_{0.1}O_2}, the theoretical magnetic moment per formula unit (f.u.) $\mu_\textrm{th} = 2.10\,\mu_\textrm{B}/f.u.$.}
\begin{tabular}{cccc}
Parameter &S1 &S2 & Sample in Ref.\cite{Wikberg2010}\\
\hline
$T_g$ (K)	&$8.0(2)$	&$8.0(2)$	&$20$ \\
$T_\textrm{ZFC-FC}$ (K)	&$64(2)$	&$122(2)$	&55 \\
$\theta_\textrm{CW}$ (K)	&$-8(1)$ 	&$-10(3)$	&$-25$ \\
$\mu_\textrm{eff}$ ($\mu_\textrm{B}/f.u.$)	&$2.26(3)$	&$2.08(2)$	&$2.07$ \\
\hline
\end{tabular}
\end{table}

In a field of 1000\,Oe, the reciprocal susceptibility $\chi^{-1}(T)$ is linear above 200\,K and was used to fit to the Curie-Weiss law, $\chi = \dfrac{C}{T-\theta_\textrm{CW}}$, where $C$ is the Curie constant and $\theta_\textrm{CW}$ is the Curie-Weiss temperature. Parameters for the Curie-Weiss fits are summarised in Table~\ref{tab:Table3}. The Curie-Weiss temperatures are negative for both samples, indicating net antiferromagnetic interactions. The calculated moment per formula unit (f.u.) for S1 of 2.26(3)\,$\mu_\textrm{B}/f.u.$ is greater than the theoretical value of 2.10\,$\mu_\textrm{B}/f.u.$, indicating a higher concentration of high spin species (\ce{Ni^{2+}} with $S = 1$ and \ce{Mn^{4+}} with $S = \sfrac{3}{2}$) as compared to the nominal stoichiometry. By contrast, the moment for S2 of 2.08(2)\,$\mu_\textrm{B}/f.u.$ is consistent with the theoretical value, indicating that S2 is almost stoichiometric.

\begin{figure}
\centering
\includegraphics[width=0.7\textwidth]{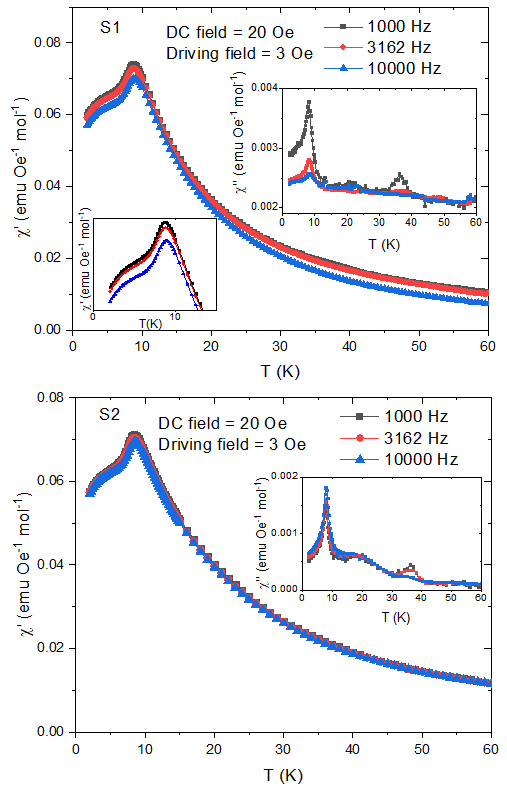}
\caption{\label{fig:Fig5} Real component of ac susceptibility $\chi^\prime(T)$ for S1 (upper panel) and S2 (lower panel) with a dc field of 20\,Oe and a driving field of 3\,Oe at different frequencies (labelled on the panels). Inset: imaginary susceptibility component $\chi^{\prime\prime}(T)$.}
\end{figure}
 
We carried out ac susceptibility measurements to probe the dynamics of the magnetic transition in both samples. The real component of the ac susceptibility, $\chi^\prime(T)$, shows a peak at 8\,K for both samples [Fig.~\ref{fig:Fig5}], consistent with the dc susceptibility measurements; it also shows an additional shoulder at $\approx 5$\,K, which could be indicative of a second transition at lower temperatures. The transitions in sample S1 shift with increasing frequency [Fig.~\ref{fig:Fig5} inset], consistent with spin-glass-like behaviour reported for other Li ion battery cathode systems \cite{Shirakami1998,Chernova2007,Wikberg2010a,Bie2013}. However, the transitions in sample S2 shows no such shift within the resolution of our measurements. The imaginary component of the ac susceptibility, $\chi^{\prime\prime}(T)$, shows additional peaks at 20\,K and 35\,K in both samples [Fig.~\ref{fig:Fig5} insets]. A previous study on Li\,NMC\,811 reported a frequency-dependent peak at $24 < T < 70$\,K, consistent with our measurements \cite{Wikberg2010}. Such additional peaks in $\chi^{\prime\prime}(T)$ at frequencies up to 1\,kHz have also been reported for other Li ion battery cathode materials such as \ce{LiNi_{0.5}Mn_{0.5}O_2} and \ce{LiNi_{0.4}Mn_{0.4}Co_{0.2}O_2}, and have been attributed to spin reorientation transitions \cite{Chernova2007}.

\subsection{\label{sec:chemcomp}Determination of chemical composition}

We now discuss refinement of the chemical composition of the two Li\,NMC\,811 samples. The refinement was subject to the following constraints: a) charge balance; b) the  the $3a$, $3b$ and $6c$ crystallographic sites (corresponding to Li, TM and O respectively) were fully occupied; c) the magnetic moment was  consistent with $\mu_\textrm{eff}$, the magnetic moment obtained from the Curie-Weiss fit [Table\,\ref{tab:Table3}]; d) the composition of the TM ions were consistent with the values obtained from elemental analysis within error. We further reduced the number of free parameters by noting that, since \ce{Co^{3+}} has $S=0$, its composition cannot be constrained using magnetometry. Hence the Co composition was fixed to the nominal value 0.1, consistent with elemental analysis. Previous neutron diffraction studies on \lno{} have examined the possibility of Li/Ni site disorder such as \ce{\{Li_{1-x}Ni_x\}_{3a}[Ni_{1-x}Li_x]_{3b}O_2}  and \ce{\{Li_{1-x}Ni_x\}_{3a}[Ni_{1-y}Li_y]_{3b}O2} and ruled it out for near-stoichiometric samples \cite{Pouillerie2001,Chung2005}. Our refinements also indicate the absence of \ce{Li^+} in the TM layers and so a single parameter $x$ was used to refine the \ce{Ni^{2+}} excess in the \ce{Li^+} layers. The composition \linmcoffstoic{} was refined for a range of ($x$, $y$) values consistent with the above constraints and the final values were chosen corresponding to the refinement with the minimum $\chi^2$. 

Both samples of Li NMC 811 are slightly Li-deficient, consistent with previous studies on \lno{}; however, their compositions are different. Sample S1 has the formula \ce{Li_{0.975(2)}Ni_{0.805(4)}Mn_{0.120(2)}Co_{0.1}O_2} while S2 has the formula \ce{Li_{0.998(2)}Ni_{0.808(4)}Mn_{0.094(2)}Co_{0.1}O_2}, corresponding to \ce{Ni^{2+}} excess in the \ce{Li^+} layers of 2.5(2)\% and 0.2(2)\%, respectively.

\subsection{Polarised neutron diffraction}

\begin{figure}
\centering
\includegraphics[width=0.7\textwidth]{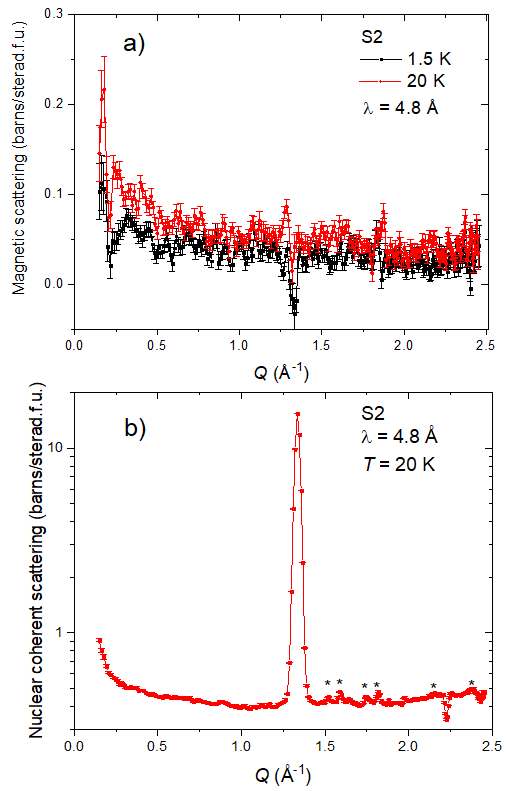}
\caption{\label{fig:Fig6} a) Magnetic scattering of S2 at 1.5\,K (black squares) and 20\,K (red diamonds). b) Nuclear coherent scattering of sample S2 at 20\,K plotted on a logarithmic scale. Impurity peaks belonging to \ce{Li_2CO_3} are marked with a *.}
\end{figure}

Neutron scattering experiments using XYZ polarisation analysis on the D7 instrument at the ILL enable separation of the nuclear coherent, nuclear-spin incoherent, and magnetic scattering contributions from the sample. Thus they are ideal for investigating diffuse scattering in disordered magnetic systems \cite{Stewart2009}. Our ac susceptibility measurements indicated a static magnetically-disordered state for S2 and so polarised neutron scattering measurements were carried out to investigate the nature of the transition at 8\,K. Figure~\ref{fig:Fig6}(a) shows the magnetic scattering as a function of momentum transfer $Q$ for S2 at 1.5\,K (below $T_g = 8$\,K) and 20\,K (above $T_g = 8$\,K). The negative intensity at $Q \approx 1.3$ \AA$^{-1}$ is an artifact from the subtraction of the nuclear Bragg peak and does not have any physical significance. No magnetic Bragg peaks are observed, consistent with the absence of long-range magnetic order; however, there is a broad diffuse feature at low $Q$ indicative of short-range spin correlations. A previous inelastic neutron scattering study on \lnooffstoic{}, $x = 0.029(1)$, reported a decrease in the inelastic channel and an increase in the elastic line on cooling through $T_g = 15$\,K, consistent with spin freezing \cite{Clancy2006}. The elastic scattering at 1.7\,K in \cite{Clancy2006} showed no magnetic Bragg peaks, only broad magnetic diffuse scattering at low $Q$. The feature observed in our magnetic scattering for S2 is qualitatively similar to this previous report on \lno{}; however, it was not possible to carry out quantitative modelling of the magnetic interactions due to the weak magnetic scattering and contributions from multiple magnetic species. 

The nuclear coherent scattering intensity at $T = 20$\,K is plotted on a logarithmic scale as a function of $Q$ in Figure~\ref{fig:Fig6}(b). Several very weak peaks from an impurity phase are observed along with the single nuclear Bragg peak from the main phase; these were identified to be from \ce{Li_2CO_3}. This \ce{Li_2CO_3} phase was beyond the detection limit of the PXRD and PND data for S2 used for our structural Rietveld refinements; however, it is observed here due to the separation of the coherent and incoherent nuclear contributions, which improves the signal-to-noise ratio in the coherent Bragg scattering. Due to the limited $Q$ range and the presence of only a single nuclear Bragg peak from the main phase, it was not possible to carry out a structural refinement to quantify the exact amount of \ce{Li_2CO_3}; however, based on our structural analysis for S1 (for which the \ce{Li_2CO_3} impurity was visible in the PND data plotted on a logarithmic scale), we can place an upper bound of 0.2(3)\,wt\%. 

\section{Discussion}
We now discuss the key features of our magnetic measurements and structural refinements on the two commercial samples of Li\,NMC\,811.

The ZFC transition temperature in our Li\,NMC\,811 samples is consistent with previous reports on \lno{} samples with similar values of $x$ in \lnooffstoic{} ($T_g = 9$\,K) \cite{Reimers1993,Clancy2006}, as well as other Ni-rich Li ion cathode materials such as \ce{LiNi_{0.8}Co_{0.15}Al_{0.05}O_2} (Li\,NCA) ($T_g = 6.5$\,K) \cite{Liu2018} and \ce{LiNi_{0.6}Mn_{0.2}Co_{0.2}O_2} (Li\,NMC\,622) ($T_g = 7.2$\,K) \cite{Dahbi2012}. This indicates that the magnetic properties are likely to be still dominated by the $S = \sfrac{1}{2}$ \ce{Ni^{3+}} and $S = 1$ \ce{Ni^{2+}} spins.  A previous study on a sample of Li\,NMC\,811 had reported $T_g \approx20$\,K and $T_\textrm{ZFC-FC} \approx 50$\,K; however, the \ce{Ni^{2+}} excess in the \ce{Li^+} layers (calculated using XRD) was 3.9\% \cite{Wikberg2010}, which is greater than both our samples, and the \ce{Mn^{4+}} composition was not refined. Previous investigations on \lnooffstoic{} have shown that even a slight change in composition can dramatically alter the transition temperature; for example, $T_g = 7.5$\,K for $x = 0.004$ and 8.6\,K for $x = 0.015$ \cite{Bianchi2001}. The transition temperature for our samples, $T_g = 8.0(2)$\,K, is consistent with lower values of $x$ (2.5(2)\% and 0.2(2)\%  for S1 and S2, respectively) as compared to the previous study. However the trend in $T_\textrm{ZFC-FC}$ differs from previous reports on \lnooffstoic{} where more off-stoichiometric samples have reported a higher value of $T_\textrm{ZFC-FC}$. Our structural analysis shows that S2 is more stoichiometric than S1 and so it is unclear why it has a higher $T_\textrm{ZFC-FC}$. One possible explanation is the presence of trace amounts of an amorphous magnetic impurity phase in S2, below the detection limit of our diffraction data, causing the irreversibility in the ZFC-FC curves at higher temperatures. 

Our magnetic measurements demonstrate that the sample dependence of the magnetic properties widely reported for the parent material \lno{} persists in Li\,NMC\,811. The origin of the sample dependence in \lno{} is the off-stoichiometry \lnooffstoic{} and $x$, the \ce{Ni^{2+}} excess in the \ce{Li^+} layers.  However, the situation is more complex in Li\,NMC\,811 due to the presence of multiple magnetic species. For ideal stoichiometry, the formula can be written as \ce{LiNi^{2+}_{0.1}Ni^{3+}_{0.7}Mn^{4+}_{0.1}Co^{3+}_{0.1}O_2}, the magnetic species being $S = \sfrac{1}{2}$ \ce{Ni^{3+}}, $S = 1$ \ce{Ni^{2+}} and $S = \sfrac{3}{2}$ \ce{Mn^{4+}}, whereas \ce{Co^{3+}} with $S = 0$ plays the role of non-magnetic site dilution in the TM layers. However, the composition for our samples is \linmcoffstoic{}, with $x = 0.025(2)$, $y = 0.120(2)$ for S1, and $x = 0.002(2)$, $y = 0.094(2)$ for S2. Thus the amount of each magnetic species present depends on the composition: ($0.9-x$) \ce{Ni^{3+}}  ($S = \sfrac{1}{2}$), ($y + 2x$) \ce{Ni^{2+}} ($S = 1$) and $y$ \ce{Mn^{4+}} ($S = \sfrac{3}{2}$). Since $x$ and $y$ are both different for S1 and S2 (S2 is closer to nominal stoichiometry), the relative number of magnetic species is also different in these samples. Additionally, $x$ \ce{Ni^{2+}} migrate to the \ce{Li^+} layers, introducing a competition between the inter-layer and intra-layer interactions dependent on $x$, analogous to \lno{}. Further, it has been observed that a higher deviation from nominal stoichiometry in \lnooffstoic{} corresponds to a greater tendency for spin-glass-like behaviour \cite{Clancy2006}. Thus S1, which is more off-stoichiometric, exhibits a frequency dependent spin-glass-like transition whereas S2, close to ideal stoichiometry, shows no such frequency dependence in the ac susceptibility.  

We find that a combination of elemental analysis, bulk magnetic measurements, and diffraction is essential to provide an accurate quantitative estimate of the composition for Li\,NMC\,811 samples with close to ideal stoichiometry. Previous studies have often set the composition to the values determined from elemental analysis and carried out structural refinements using X-ray diffraction data only \cite{Dahbi2012,Mauger2012,Zhang2010}. Our results indicate that this approach may need to be treated with caution for Ni-rich compositions, which tend to be Li-deficient. Elemental analysis provides the average composition value for each element; that is, the contributions from the main phase (Li\,NMC\,811) as well as any impurity phases (\ce{Li_2CO_3}) are both included. PXRD is much less sensitive to the presence of light elements like Li, C, and O, and so no \ce{Li_2CO_3} impurity Bragg peaks are visible in the room temperature PXRD pattern. By contrast, PND is much more sensitive to the presence of these elements; our room temperature PND data for S1 and polarised neutron diffraction data at 20\,K for S2 provide conclusive evidence for the presence of \ce{Li_2CO_3}. As the Li composition from elemental analysis was close to 1 and the neutron data confirms the presence of \ce{Li_2CO_3}, we can conclude that the Li\,NMC\,811 samples are indeed Li-deficient. This is consistent with our combined structural refinements. Recent high resolution powder diffraction studies on Li\,NMC oxides have also indicated the necessity of using PXRD and PND data to determine the stoichiometry accurately \cite{Yin2020,Liu2017}. By including an additional constraint on the total magnetic moment from our magnetic measurements, we are able to increase the accuracy of our refined compositions for Li, Ni, and Mn in Li\,NMC\,811. Techniques such as Li nuclear magnetic resonance (NMR) could also be used to quantify the Li composition in Li\,NMC\,811 more accurately as the signal would be well separated for paramagnetic (Li\,NMC\,811) and diamagnetic (\ce{Li_2CO_3}) Li-containing phases.

Measurements comparing the electrochemical performance of S1 and S2, Figure S1, suggest that the degree of \ce{Ni^{2+}} excess in \ce{Li^+} layers ( = 2.5(2)\% for S1 and 0.2(2)\% for S2 respectively) has no significant influence on the rate performance. This is a significant departure from previous results on \lno{} where the presence of even small amounts of \ce{Ni^{2+}} excess in \ce{Li^+} layers in \lno{} has been repeatedly linked to deterioration in cycling performance as the \ce{Ni^{2+}} significantly hinders \ce{Li^+} mobility \cite{Bianchini2019}. It is possible that the additional structural stability of Li\,NMC\,811 makes the electrochemical performance more robust to low levels of off-stoichiometry (at least up to 2 - 3\% of \ce{Ni^{2+}} excess in \ce{Li^+} layers). Our results indicate that magnetic measurements are a convenient tool to identify such `good quality' samples of Li\,NMC\,811 and other Ni-rich Li ion TM oxide battery cathode materials (which are prone to off-stoichiometry and migration of \ce{Ni^{2+}} into the \ce{Li^+} layers) prior to carrying out long-term electrochemical tests.

\section{Conclusion}

We have carried out elemental analysis, room temperature X-ray and neutron diffraction, bulk magnetic measurements, and polarised neutron scattering measurements on two powder samples of Li\,NMC\,811 from a commercial supplier. Our combined PXRD and PND structural refinements using constraints from all these techniques show that the samples have the composition \ce{Li_{0.975(2)}Ni_{0.805(4)}Mn_{0.120(2)}Co_{0.1}O_2} for S1 and  \ce{Li_{0.998(2)}Ni_{0.808(4)}Mn_{0.094(2)}Co_{0.1}O_2} for S2 respectively. Magnetic measurements reveal a transition at 8\,K for both samples, but the ZFC-FC curves deviate at very different temperatures: $T_\textrm{ZFC-FC}$ = 64\,K for S1 and 122\,K for S2. However, based on the two samples studied here, a higher $T_\textrm{ZFC-FC}$ does necessarily indicate a less stoichiometric sample. The nature of the transition is also different: shifting with increase in frequency for S1 while no such shift is observed for S2. This latter point is attributed to the fact that S1 is more off-stoichiometric and so shows a greater tendency for spin freezing, analogous with the parent compound \lnooffstoic{}. 

Thus a combination of experimental techniques such as elemental analysis, diffraction, and bulk magnetic measurements are required to accurately determine the structure and chemical composition of Li\,NMC\,811 and other Ni-rich Li ion battery cathode materials with close to ideal stoichiometry. Our results indicate that ac susceptibility measurements could be investigated as a sensitive technique to identify off-stoichiometry in these materials.
\begin{acknowledgement}

We thank the Science and Technology Facilities Council (STFC) for provision of ISIS Xpress Access beam time on GEM and the Institut Laue-Langevin for allocation of EASY beam time on D7. P.M., C.X., Z.R., C.P.G. and S.E.D. acknowledge funding support from the Faraday Institution EPSRC Grant EP/S003053/1. Magnetic measurements were carried out using the Advanced Materials Characterisation Suite, funded by EPSRC Strategic Equipment Grant EP/M000524/1. We thank Cheng Liu for support with the MPMS and PPMS equipment. We thank Craig A. Bridges (ORNL) and Binod Rai (ORNL) for useful feedback. J.A.M.P.'s work at Cambridge (contribution to neutron data reduction) was supported by Churchill College, University of Cambridge. J.A.M.P.'s work at ORNL was supported by the Laboratory Directed Research and Development Program of Oak Ridge National Laboratory, managed by UT- Battelle, LLC for the US Department of Energy (contribution to data analysis). This manuscript has been authored by UT-Battelle, LLC under Contract No. DE-AC05-00OR22725 with the U.S. Department of Energy.  The United States Government retains and the publisher, by accepting the article for publication, acknowledges that the United States Government retains a non-exclusive, paid-up, irrevocable, world-wide license to publish or reproduce the published form of this manuscript, or allow others to do so, for United States Government purposes.  The Department of Energy will provide public access to these results of federally sponsored research in accordance with the DOE Public Access Plan (http://energy.gov/downloads/doe-public-access-plan).
\\ Supporting data can be found at https://doi.org/10.17863/CAM.62029. 
\\ Neutron scattering data can also be found at ISIS dois: http://doi.org/10.5286/ISIS.E.RB1890370-1, http://doi.org/10.5286/ISIS.E.101124089, and ILL doi: https://doi.ill.fr/10.5291/ILL-DATA.EASY-432.

\end{acknowledgement}




\bibliography{achemso-demo}

\end{document}


\section{\label{sec:ICPerr}ICP-OES error analysis}

To understand the error in the ICP-OES measurement, error calculations were made from the two largest sources of uncertainty: the error associated with the sample measurement and the error in the linear calibration.  From these values, error calculations were performed at the 95\% confidence level as discussed in \cite{Miller2018}. The standard deviation of the sample measurement is simply the standard deviation of the three replicate measurements and represents how repeatable the individual measurement is for a given sample. The prediction interval, $s(x_0)$, represents how accurate the instrument response is based on a linear relationship between the sample concentration and intensity at a given wavelength.  The prediction interval can be calculated using equation \ref{eq:A1}.

\begin{equation} \label{eq:A1}
s(x_0) = \frac{RSD}{b}\sqrt{\frac{1}{N} + \frac{1}{n} + \frac{(\bar{y_0}-\bar{y})^{2}}{b^2\Sigma_{i=1}^{n}(x_i - \bar{x})^2}}
\end{equation}
\\where 
\\RSD = residual standard deviation of $y$ with $x$
\\$n$ = number of calibration points = 4
\\$N$ = number of repeat calibration points / replicates = 3
\\$b$ = slope of linear calibration 
\\$\bar{y_0}$ = mean of $N$ measurements of $y$-value (intensities) for the sample
\\$\bar{y}$ = mean of the $y$-values of the calibration standards
\\$x_i$ = $x$-value (concentration) of the standards
\\$\bar{x}$ = mean of the $x_i$ values of the samples

To combine the error at each wavelength at the confidence level, the confidence interval was calculated for both the standard deviation of the sample measurement ($\mu_{\lambda_n,s} = t_{N-1}\frac{StDev_{\lambda_s}}{\sqrt{N}}$) and for the prediction interval ($\mu_{\lambda_n,s(x_0)} = t_{n-2}s(x_0)$) using a two sided t-statistic.  In our case, both t-statistics have 2 degrees of freedom ($t_{N-1} = t_{n-2} = 4.3$).  The standard deviation of the sample measurement was then combined with the prediction interval to obtain the total error of the measurement at a given wavelength $\mu_\lambda = \sqrt{{\mu_{\lambda_n,s}}^2 + {\mu_{\lambda_n,s(x_0)}}^2}$). The errors are given in Table \ref{tab:tableS1}. Since the concentrations of each element were measured at different wavelengths (2 for Co, 3 for Mn and 4 for Ni) and averaged to obtain the mean concentration in the measurement, the confidence interval for each wavelength was combined using $\mu_{element,k} = \frac{{\bar{\mu}}_j}{\sqrt{No. of \lambda}}$.   Finally, the confidence interval for each element was calculated using the expression
$\mu_{element,composition} = \sqrt{\sqrt{{\mu_{Ni,k}}^2 +{\mu_{Mn,k}}^2 +{\mu_{Co,k}}^2} + {\mu_{element,k}}^2}$ 
since the transition metals were assume to have a total fraction of 1 and the Li composition was calculated by dividing the number of moles of lithium by the number of moles of transition metal.\\ 

\begin{table}
\caption{\label{tab:tableS1}ICP-OES error analysis for samples S1 and S2.}
\begin{adjustbox}{max width=\textwidth}
\begin{tabular}{ccccccccccc}
\hline
Element &Li &Ni &Ni &Ni &Ni &Mn &Mn &Mn &Co &Co \\ 
Wavelength (nm) &610.362 &216.556 &221.647  &230.3 & 231.6 &257.61	&259.373 &260.569 &228.616	&237.862 \\ \hline
\multicolumn{11}{c}{S1}\\ \hline
Mean &1.057 &0.802 &0.796 &0.796 &0.795 &0.099	&0.100	&0.102 &0.102 &0.103\\
StDev (sample)(\%) &1.7 &0.4 &0.5 &0.5 &0.5 &1.1 &1.0 &1.1 &0.5 &1.0\\
$\mu_{\lambda,s}$ (\%) &4.3 &0.9 &1.2 &1.2 &1.2 &2.7 &2.5 &2.7 &1.2 &2.6\\
$\mu_{\lambda,s(x_0)}$ (\%) &0.5 &0.7 &0.9 &2.8 &0.9 &3.1 &2.6 &2.6 &2.8 &3.4\\
$\mu_\lambda$ (\%) &2.1 &2.9 &3.7 &12.2 &3.7 &13.4 &11.4 &11.0 &12.2 &14.8\\ \hline
\multicolumn{11}{c}{S2}\\ \hline
Mean &1.007 &0.803 &0.800 &0.798 &0.797 &0.098	&0.099	&0.100 &0.102 &0.101\\
StDev (sample)(\%) &2.2 &0.2 &0.2 &0.1 &0.1 &1.4 &1.6 &1.6 &0.2 &2.1\\
$\mu_{\lambda,s}$ (\%) &5.4 &0.5 &0.4 &0.3 &0.2 &3.4 &3.9 &4.1 &0.6 &5.3\\
$\mu_{\lambda,s(x_0)}$ (\%) &0.6 &0.8 &1.0 &0.6 &1.0 &4.1 &3.5 &3.3 &3.7 &4.6\\
$\mu_\lambda$ (\%) &2.7 &3.4 &4.4 &2.8 &4.4 &17.6 &14.8 &14.4 &15.9 &19.7\\ 
\hline
\end{tabular}
\end{adjustbox}
\end{table}

\section{\label{sec:echem}Electrochemical performance}
\begin{figure}
\centering
\includegraphics[width=0.7\textwidth]{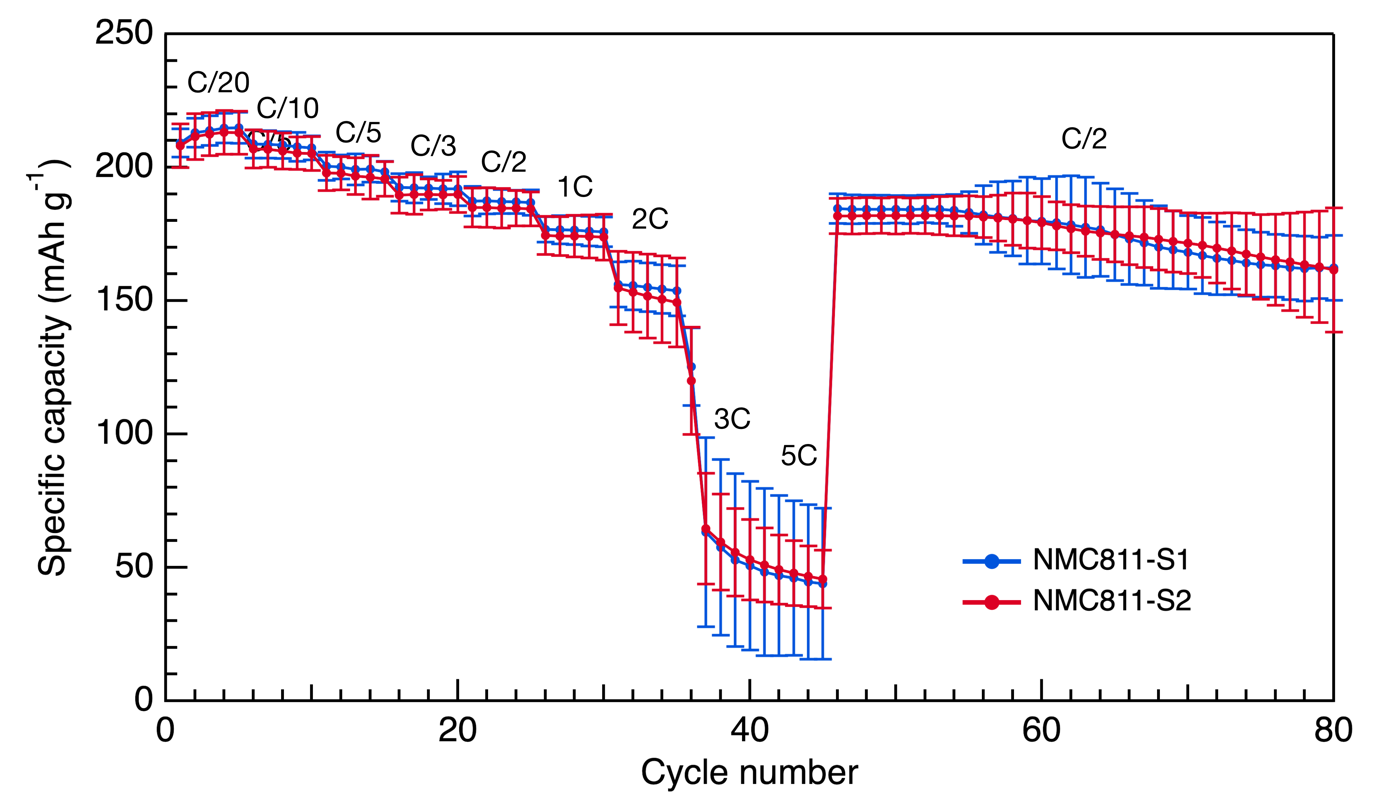}
\caption{\label{fig:FigS1} Discharge capacities of S1 and S2 at various cycling rates between 3.0 V and 4.4 V vs. Li. The capacity is normalized to the mass of Li NMC 811, and the C-rate is calculated based on a reversible capacity of 200 mAh g$^{-1}$. The error bars are calculated based on three cells for each sample.}
\end{figure}

We evaluated the electrochemical performance of S1 and S2 at various charge/ discharge rates as the \ce{Ni^{2+}} excess in the \ce{Li^+} layers is expected to influence the Li ion diffusivity, and therefore impact the rate capability of the cathode material \cite{Makimura2016}. The experiments were carried out in half-cell configuration, i.e. with Li metal as the anode, and three cells per sample were tested. The average discharge capacities at various rates are show in Figure \ref{fig:FigS1}. 

Both samples show a discharge capacity of ~ 210 mAh g$^{-1}$ at C/20 rate, which is in good agreement with literature that Li NMC 811 cathodes typically show capacities above 200 mAh g$^{-1}$ at slow rates \cite{Ryu2018}. S1 and S2 exhibit good rate capability, with no major capacity decreases until extremely high rates (i.e. 3C and 5C) and the capacity-rate profiles of the two samples are also very similar. However, the electrochemical performance may show differences if they are cycled at higher voltages $\geq 4.5$ V where there are additional degradation mechanisms including greater \ce{Ni^{2+}} migration into the \ce{Li^+} layers.

\bibliography{achemso-demo}